\documentclass[prl,superscriptaddress,amsmath,amssymb,showpacs,noshowkeys,a4paper,twocolumn]{revtex4-1}
\usepackage{graphicx}% Include figure files
\usepackage{dcolumn}% Align table columns on decimal point
\usepackage{bm}% bold math
\usepackage[usenames]{color}
\usepackage [babel=true]{csquotes}
\usepackage{MnSymbol}
\usepackage{epstopdf}
\definecolor{mygray}{gray}{0.5}
\usepackage{enumerate}
\usepackage{gensymb}
\usepackage{color}
\newcommand{\masquer}[1]{}

\newcommand{\affmsc}{\affiliation{Mati\`ere et Syst\`emes Complexes, CNRS and Universit\'e Paris Diderot UMR 7057, B\^atiment Condorcet, 10 rue Alice Domon et L\'eonie Duquet, 75013 Paris, France}}

\begin{document}

\title{{Constant} \masquer{Universal}Froude number in a circular hydraulic jump and its implication on the jump radius selection.}

\author{A.~Duchesne}
\affmsc

\author{L.~Lebon}
\affmsc

\author{L.~Limat}
\affmsc

\begin{abstract}
\masquer{As well known, t} {T}he properties of a standard hydraulic jump depend critically on a Froude number $Fr$ defined as the ratio of the flow velocity to the gravity waves speed. In the case of a horizontal circular jump, the question of the Froude number is \masquer{still}not well \masquer{understood}{documented}. {Our experiments show that $Fr$ measured just after the jump is locked on a constant value that does not depend on flow rate $Q$, kinematic viscosity $\nu$ and surface tension $\gamma$.}\masquer{We have investigated quantitatively this question and a universal Froude number was observed just after the jump.} Combining this result to a lubrication description of the outer flow yields, \masquer{under certain conditions, to}{under appropriate conditions,} a new and simple law \masquer{linking the flow rate $Q$ to}{ruling} the jump radius $R_J$: $R_J (ln (\frac{R_\infty}{R_J}))^{3/8} \sim Q^{5/8}\nu ^{-3/8}${, in excellent agreement with our experimental data. This unexpected result asks an unsolved question to all available models}.

\end{abstract}

\maketitle
Structure formations in free surface flows remain a major source of complexity in hydrodynamics. Perhaps the most well known example is the \enquote{hydraulic jump}, in which one observes a sudden transition from a high speed, supercritical, open channel flow to a subcritical one, with a sudden jump of the fluid depth \cite{chanson_2009, Lighthill_78}. This phenomenon is ubiquitous and can be observed at very different scales: dam release flows \cite{chanson_2009}, tidal bores on rivers \cite{chanson_mascaret} and kitchen sinks when a vertical jet of liquid hits a horizontal surface.  The circular liquid wall observed in this very last case has even motivated recent model experiments for astrophysics \cite{Rousseaux_2011}, to mimic the competition between the speed of a wave and that of a radial flow (white hole equivalents) \cite{Weinfurtner_2011}. The natural dimensionless number used to describe this competition is the Froude number $Fr$ defined \cite{belanger} as the ratio of the flow velocity to the gravity waves speed $Fr=U/\sqrt{gh}$ where $U$ is the average flow velocity, $h$ the fluid thickness and $g$ the acceleration of gravity. \masquer{It is known that}{In general,} the properties of a standard hydraulic jump depends critically on the $Fr$ value \cite{chanson_2009}, that is always larger than one upstream and smaller than one downstream, the two values being dependent on each other via the mass and \masquer{impulsion}{momentum}  conservation laws written at the jump location.\\
\indent As\masquer{is} well known, several complex physical effects are mixed, even in \masquer{this}{the} reasonably simple case  {of a circular hydraulic jump at moderate flow rate}: on one hand, the jump can be understood as a shock front for surface gravity waves \cite{Lighthill_78}, but its formation can also be understood as something similar to the growth and detachment of a boundary layer close to the solid substrate \cite{ Bohr_96}, with the formation of toroidal recirculations all around the front. In turn, this association between a front and a toroidal vortex can become unstable leading to surprising facetting  effects \cite{Bush_2006, bohr_12}.\\
	\begin{figure}[b!]
 \centering
    \begin{tabular}{cc}
      \includegraphics[width=0.5 \columnwidth]{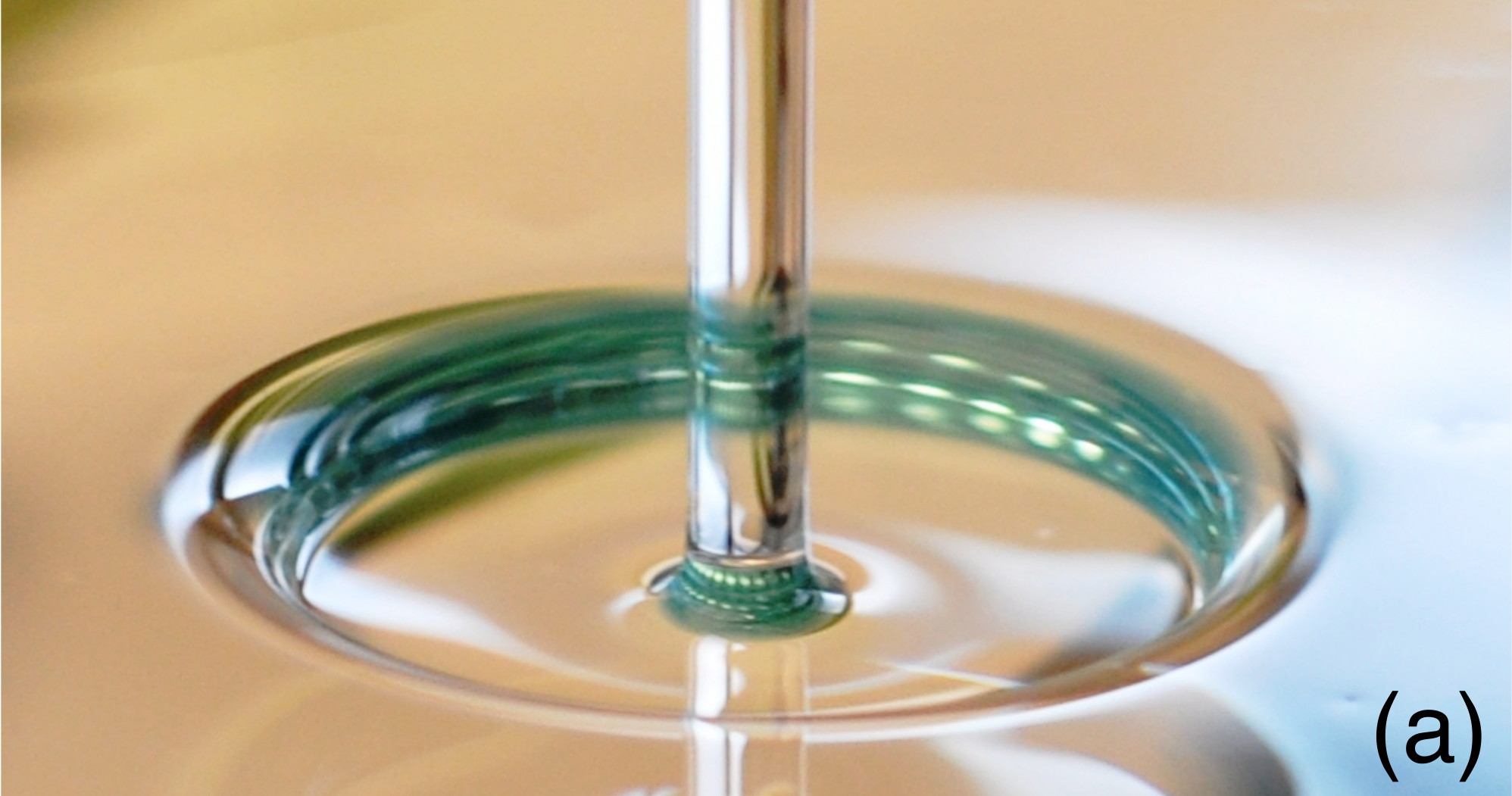}  \\
      \includegraphics[width=0.8 \columnwidth]{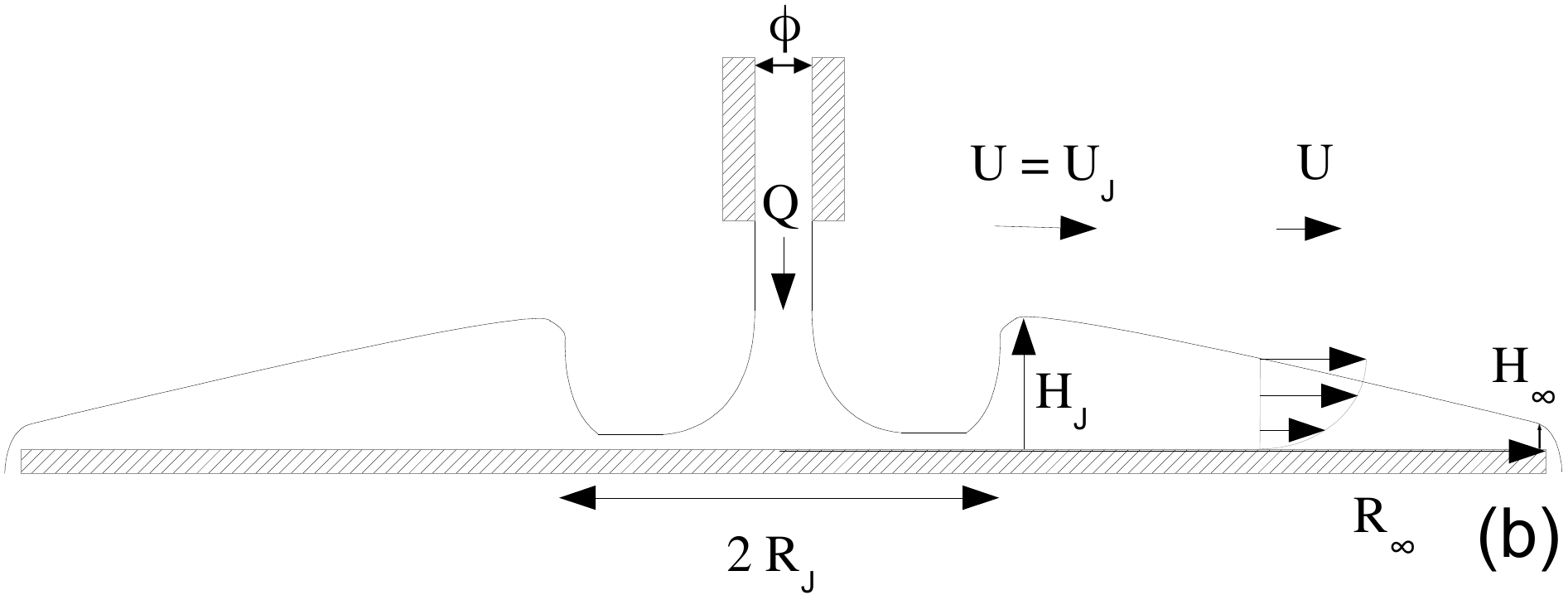}      
     \end{tabular}
\caption{\label{fig1} (a) A jet of liquid (silicone oil ($20$ cS)) impacts vertically a horizontal surface. A circular hydraulic jump is observed.
(b)  Sketch of the experimental situation. A jet of liquid issued from a vertical tube of internal diameter \masquer{$\phi=3$} {$\phi=3.2$} mm hits the center of a horizontal disk  of radius $R_\infty=15$ cm placed $4$ cm below the outlet. A circular hydraulic jump of radius $R_J$ is observed. At the exit of the jump the height is $H_J$ and the average speed $U_J$. There is no confinement wall on the disk perimeter, where $H_\infty$ is the liquid thickness.}
\end{figure}
\indent In the present letter, we show that even \masquer{the}{this} simplest form of hydraulic jump \masquer{, i.e. the circular hydraulic jump formed from an impacting jet at moderate flow rate,}  involves an unsuspected selection mechanism that fixes the Froude number value at the jump exit. This discovery has been missed, up to now, despite numerous studies \cite{Rayleigh_1914, Tani_48, Watson_JFM, Craik_81, Bohr_1993,higuera_1994, Bohr_96, Bush_2003, Bush_2006, yokoi2000, Stone_2009, dasgupta2010, Maynes_2011, rojas_2013}, the question of the precise value that this number can take having remained unaddressed for a circular hydraulic jump. Most of \masquer{the} available studies have focused rather on the question of the radius selection of the jump $R_J$, for which two main theories are available \cite{Watson_JFM, Bohr_1993}{, among a rich literature trying to improve these ones or to propose alternatives \cite{higuera_1994, Bush_2003, bohr_2003, Bush_2006, Kasimov_2008, dasgupta2010, rojas_2013}}. \\
	\indent We {here} investigate quantitatively the Froude number selection \masquer{question} with experiments performed in the case of a liquid jet impacting a horizontal disk, with no confinement walls imposing the outer thickness. The Froude number $Fr$ is accurately measured in a large range of flow rate $Q$ and appears to be constant and independent of $Q$, of kinematic viscosity $\nu$ and of surface tension $\gamma$. Surprisingly this result cannot be recovered from the two main theories of Watson \cite{Watson_JFM} and Bohr \cite{Bohr_1993} \masquer{which describe the inner flow, the flow before the jump}. \\
	\indent Furthermore, combining this result with a simple but accurate description of the outer flow (i.e. the thickness distribution of liquid), we obtain a new and simple law linking $R_J$ and $Q$ which is in excellent agreement with our data {(see Eq.~(\ref{e.4}), below)}. This law is very close to the \masquer{one} {scaling} proposed by Bohr \textit{et al.} \cite{Bohr_1993}, but involves an additional logarithmic dependence upon $R_J$ \masquer{that has been missed up to now} {that can not be neglected. These results are consistent with several developments of Bohr \textit{et al.} theory \cite{Bohr_1993, bohr_2003}, in which logarithmic effects were identified in the outer thickness distribution and in the numerical resolution of $R_J(Q)$. However, it is the first time that a so simple, analytical, law is proposed for $R_J(Q)$ and proved experimentally}.\\
	 \indent Our experiment is depicted in Fig.~\ref{fig1}.  A jet of liquid issues from a vertical tube of internal diameter \masquer{$\phi=3$} {$\phi=3.2$} mm, hits the center of a transparent glass disk of radius $R_\infty=15$ cm, placed $4$ cm below the outlet. The disk horizontality is tuned by using micrometer screws and observing a spirit level. Boundary conditions are key parameters for such an experiment, as experimentally studied in ~\cite{Bohr_96}. In our set-up we chose the simplest possible case: the absence of a wall fixing the outer height of the jump (see Fig.~\ref{fig1} (b)); the flow being let free to adjust this height by itself. Fixing this boundary condition and following surface tracers, we observed that the hydraulic jump was of type I  ~\cite{Bohr_96} (i.e., unidirectional surface flow {with no vortex reversing the flow at the free surface}). The flow was imposed by a gear pump in order to minimize flow rate pulsations. The range of accessible flow rates and the liquid viscosities chosen maintain the hydraulic jump in a steady laminar state \cite{Bush_2006} (no turbulence or instabilities of the jump).\\
	 \indent The flow rates are measured by using a flow meter calibrated by weighing \masquer{each}{the} liquid; the accessible range of flow rates (dependent on the liquid viscosity) is typically 5-60 $\pm$ 0.25 $\mathrm{cm^3.s^{-1}}$. We measured the hydraulic jump radii by visualizing from below, through the glass plate. The obtained values are known with an accuracy of $\pm 0.2$ mm.\\
	 \indent Experiments were conducted with two different types of liquids: silicone oils and water-glycerine mixtures. Silicone oil has a smaller surface tension ($\sim 20$ $\mathrm{mN.m^{-1}}$), and a density close to that of water (between 0.95 and 0.965 at 25\degree C for the different silicone oils we used). We used three different kinematic viscosities: $20.4 \pm 0.6$, $44.9 \pm 1.5$ and $98.8 \pm 3$ cS. Water-glycerine mixtures were also used, with a larger surface tension (close to $65$ $\mathrm{mN.m^{-1}}$) and a density around $1.2$ ($1.19$ for the lower viscosity and $1.22$ for the higher one). The kinematic viscosities chosen are: $18 \pm 0.7$ and $44 \pm 1.5$ cS. This range of parameter allowed us to investigate the influence on the critical $Fr$ of the three main possible parameters: kinematic viscosity, surface tension and flow rate.\\ 
	\indent In order to determine the Froude number at the exit of the jump and the outer liquid thickness distribution, we used a vernier height gauge having a needle pointer. A vertical needle is put into contact with the liquid free surface, and then with the disk surface, the difference of heights giving the liquid thickness. Examples of the obtained results are shown in Fig.~\ref{fig2}. \\
			\begin{figure}[b!]
		 \centering
    \begin{tabular}{cc}
      \includegraphics[width=0.5 \columnwidth]{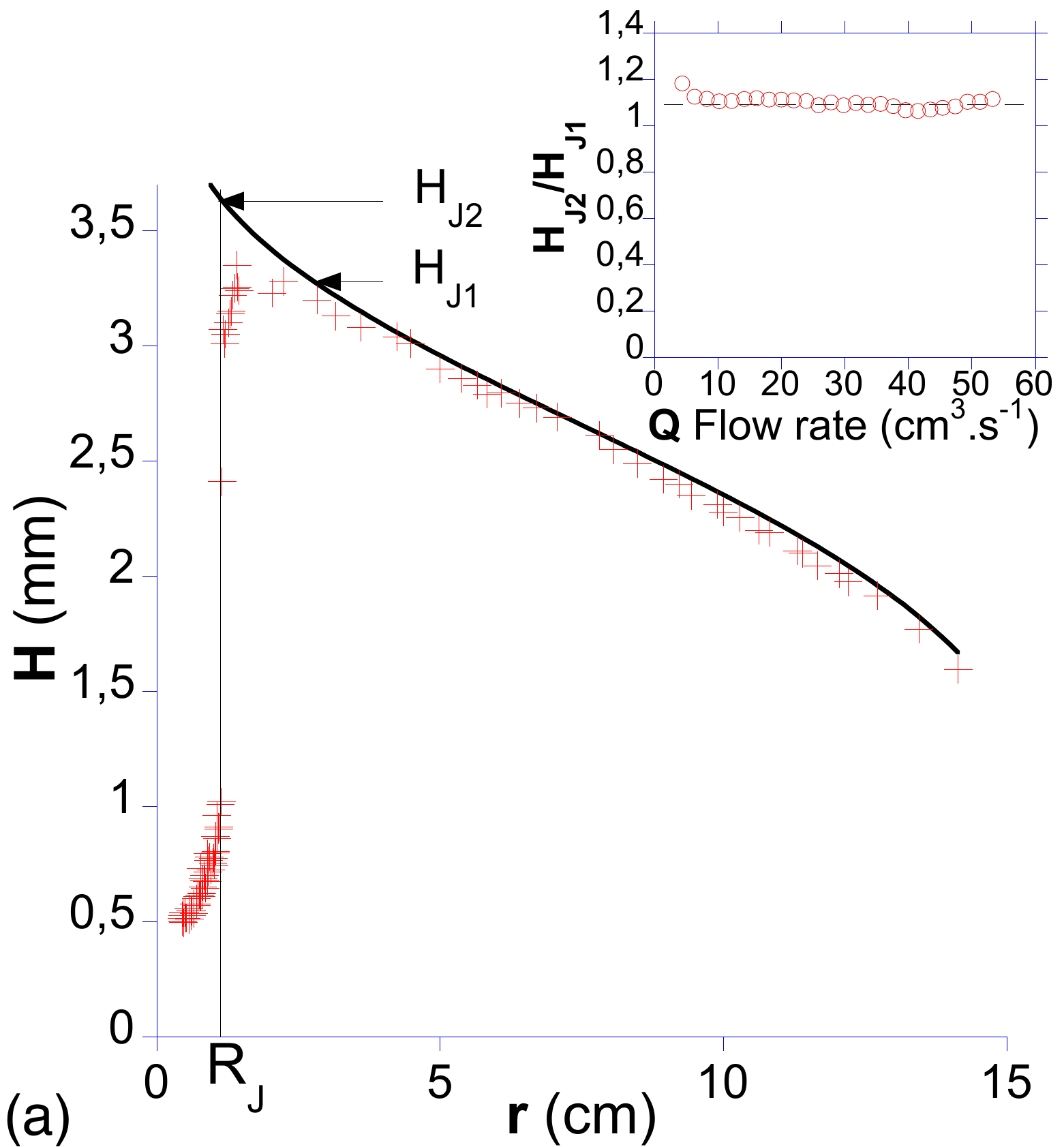}  
      \includegraphics[width=0.5 \columnwidth]{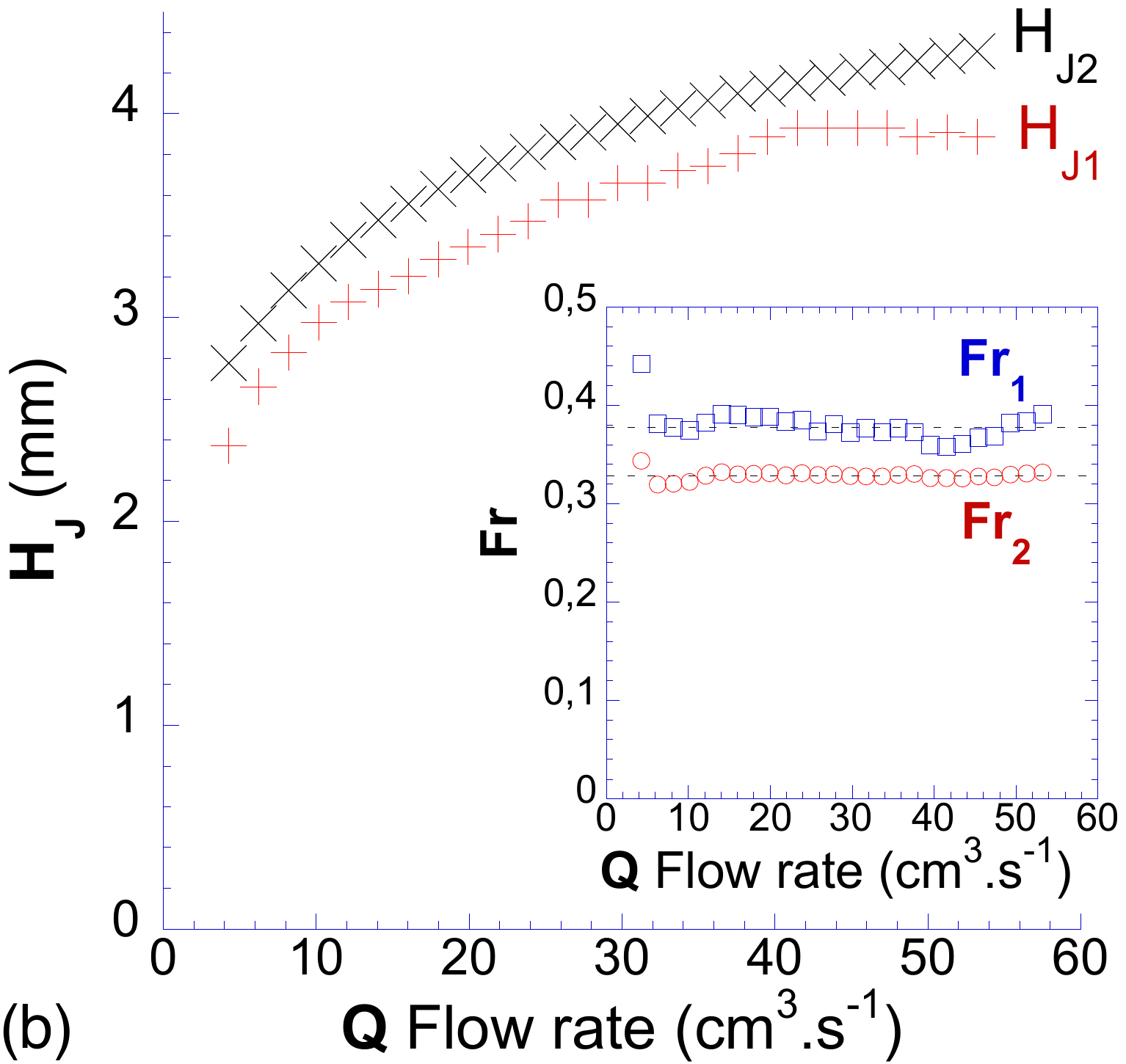}      
     \end{tabular}
     \caption{\label{fig2}(a) Liquid height profile. The liquid used is a silicone oil ($20$ cS). The measured flow rate $Q$ is $17$ $\mathrm{cm^3.s^{-1}}$ and the measured thickness at the disk position $H_\infty =1.4$ mm. Eq.~(\ref{e.3})   is plotted with a continuous line. Two values for $H_J$ can be extracted: $H_{J1}$ the measured height just at the jump exit and $H_{J2}$ the extrapolated value of the liquid thickness calculated from Eq.~(\ref{e.3}) at the jump radius $R_J$. Insert: ratio of the two possible definition of $H_J$. This ratio appears to be constant and close to $1.1$.(b) Outer height $H_J$ as a function of the flow rate $Q$ for silicone oil $20$ cS. Insert: critical Froude number observed at the jump exit with the two possible definitions on Fig.~\ref{fig2}-a.}
\end{figure}
	\indent As one can see, the {thickness distribution allows one to propose two possible definitions of the liquid thickness at the jump exit $H_J$. The first one is just the value measured $H_{J1}$, while the second is the extrapolated value $H_{J2}$ of a simple lubrication model of the outer flow described below (see Eq.~(\ref{e.3})). Both definitions allow to define a Froude number at the jump exit that reads, from mass conservation, $Fr=U_J /\sqrt{gH_J}=Q/(2\pi R_J g ^{1/2} H_J ^{3/2})$ where $R_J$ is the jump radius. The first definition is more natural, but the second will allow us to develop more easily analytical calculations. } \masquer{non-zero thickness of the front and the matching between the inner and the outer zones allow one to propose two definitions of the liquid thickness at the jump exit $H_J$.  $H_{J1}$ the height we can measure directly just at the jump exit and $H_{J2}$ the extrapolate value of the liquid thickness at the jump radius $R_J$ calculated from a simple lubrication model described below (see Eq.~(\ref{e.3})). \\
	\indent Many physical reasons may be adduced to explain this problem of matching : in our model, the presence of a toroidal vortex near the wall \cite{Bohr_96} is not taken into account, and one can also suggest that the half-poiseuille profile may not be immediately established at the exit of the jump.  But fortunately, careful measurements (see Fig.~\ref{fig2}(a) : insert) show that $H_{J1}=\alpha H_{J2}$ with $\alpha\approx 0.9$ and that this ratio $\alpha$ does not change with the flow rate or with the viscosity.\\
	\indent The conservation of mass can be written as:  $Q=2\pi r U H$, where r is the radial coordinate. At the jump position ($r=R_J$ and $H=H_J$), where $R_J$ and $Q$ can be accurately measured, this law gives $U_J$ and so, finally, $Fr=U_J /\sqrt{gH_J}$.}\\
	\indent Surprisingly a constant Froude number $Fr$ independent of the flow rate $Q$ is found (see Fig.~\ref{fig2}(b) : insert): 
	\begin{equation}
Fr_2\sim0.33 \pm 0.01,
\end{equation}
with the second definition, while a constant value is also obtained with the first one ($Fr_1\sim 0.38 \pm 0.02$). {This second result is consistent with the fact that, experimentally, $H_{J1}$ and $H_{J2}$ were found to be proportional (see Fig.~\ref{fig2}(a) : insert), i.e. $H_{J2}=\alpha H_{J1}$ with $\alpha\approx 1.09 \pm 0.03$. Note that the constancy of the Froude number and the momentum conservation impose the previous relation between the two heights with $\alpha$ depending only on the Froude numbers.}\\
\indent For both definitions, these values are smaller than one, as expected from the theory for a Froude number in the subcritical zone. But a Froude number fixed to a constant value appears to be unexpected and is not predicted by the \masquer{existing} {two main} hydraulic jump  theories :\\
	\indent (i) {In its simplest form, Bohr \textit{et al.} theory leads to a scaling  $U_J \sim Q^{1/8}\nu ^{1/8} g^{3/8} [4 ln (R_S/R_J)]^{1/4}$, where $R_S$ is the radius where the singularity happened (typically the end of the plate $R_\infty$), and a outer height $H_J\sim Q^{1/4}\nu ^{1/4} g^{-1/4} [4 ln (R_S/R_J)]^{-1/4} $, which would lead to a Froude number: $Fr_{Bohr}\sim [4 ln(R_S/R_J)]^{3/8}$. In this approximation, $Fr$ would depend on the flow rate through $R_J$, which is in contradiction with our study. A more complete version of this theory is available in refs \cite{Bohr_1993, bohr_2003} but is solved numerically with input constants ($r_0 , h_0$), which have to be measured. The authors did not checked what happened to $Fr$.} \masquer{For the outer speed at the jump position, Bohr \masquer{theory} anticipates a law with the form: $U_J \sim Q^{1/8}\nu ^{1/8} g^{3/8} [4 ln (R_S/R_J)]^{1/4}$, where $R_S$ is the radius where the singularity happened (typically the end of the plate $R_\infty$), and a outer height $H_J\sim Q^{1/4}\nu ^{1/4} g^{-1/4} [4 ln (R_S/R_J)]^{-1/4} $. So finally, Froude number in Bohr \masquer{theory} is expected to scale as: $Fr_{Bohr}\sim [4 ln(R_S/R_J)]^{3/8}$ which does depend on flow rate via $R_J$. A dependence of $R_J$ on $Q^{5/8}$ as predicted by the Bohr \masquer{theory} is tested on Fig.~\ref{fig3}(a) : insert. The agreement is quite poor and leads us to believe that a correction is missing...} \\
	\indent (ii) In the case of Bush-Watson theory the outer height is not fixed and has to be experimentally measured to determine the jump radius. So this theory lets a free parameter available and cannot predict our outer Froude number.\\
	\indent  {Most of available theories mainly target the internal flow possibly coupled to an outer flow. In contrast with these previous works, we propose here an empirical approach combining the knowledge of the outer height profile with the constant Froude number that we have found at the jump exit.}\masquer{The common point of these theories is to start the study from the inner flow profile. In contrast with these previous works, we propose here an approach built from the outer height profile.} First, the boundary conditions (no slip condition on the solid and free slip at the free interface) allow us to identify the outer flow to a parabolic profile flow governed by a balance between hydrostatic pressure and viscous friction. More precisely the equation governing the flow reads as follows: 
		\begin{equation}
		\label{e.2}
U(r) \simeq - \frac{H(r)^2}{3\nu}g\frac{dH(r)}{dr},
\end{equation}
combining this equation with the mass conservation $Q=2\pi rUH$ leads to the following solution:
		\begin{equation}
		\label{e.3}
H(r)=(H_\infty^4+\frac{6}{\pi} \frac{\nu Q}{g} ln (\frac{R_\infty}{r}))^{1/4},
\end{equation}
	where $H_\infty$ is the liquid thickness at the disk perimeter (\enquote {end} of the flow for $r=R_\infty$). {We have to notice that a similar solution has already be found by Bohr \textit{et al.} \cite{Bohr_1993,bohr_2003} and Rojas \textit{et al.} \cite{ rojas_2013} but not combined, as here, with a constant Froude number.}\\
	\indent This result can, of course, be experimentally tested: for a given flow rate  we accurately measured the outer liquid thickness profile $H(r)$ with our needle pointer. Results are shown on Fig.~\ref{fig2}(a), where the continuous line represents Eq.~(\ref{e.3}), plotted with no adjustable parameters. We see an excellent agreement with the data. \\
\indent At this point, we have to distinguish two cases depending on the wetting conditions on the glass: \\
 \begin{figure}[t!]
 		 \centering
    \begin{tabular}{cc}
      \includegraphics[width=0.75 \columnwidth]{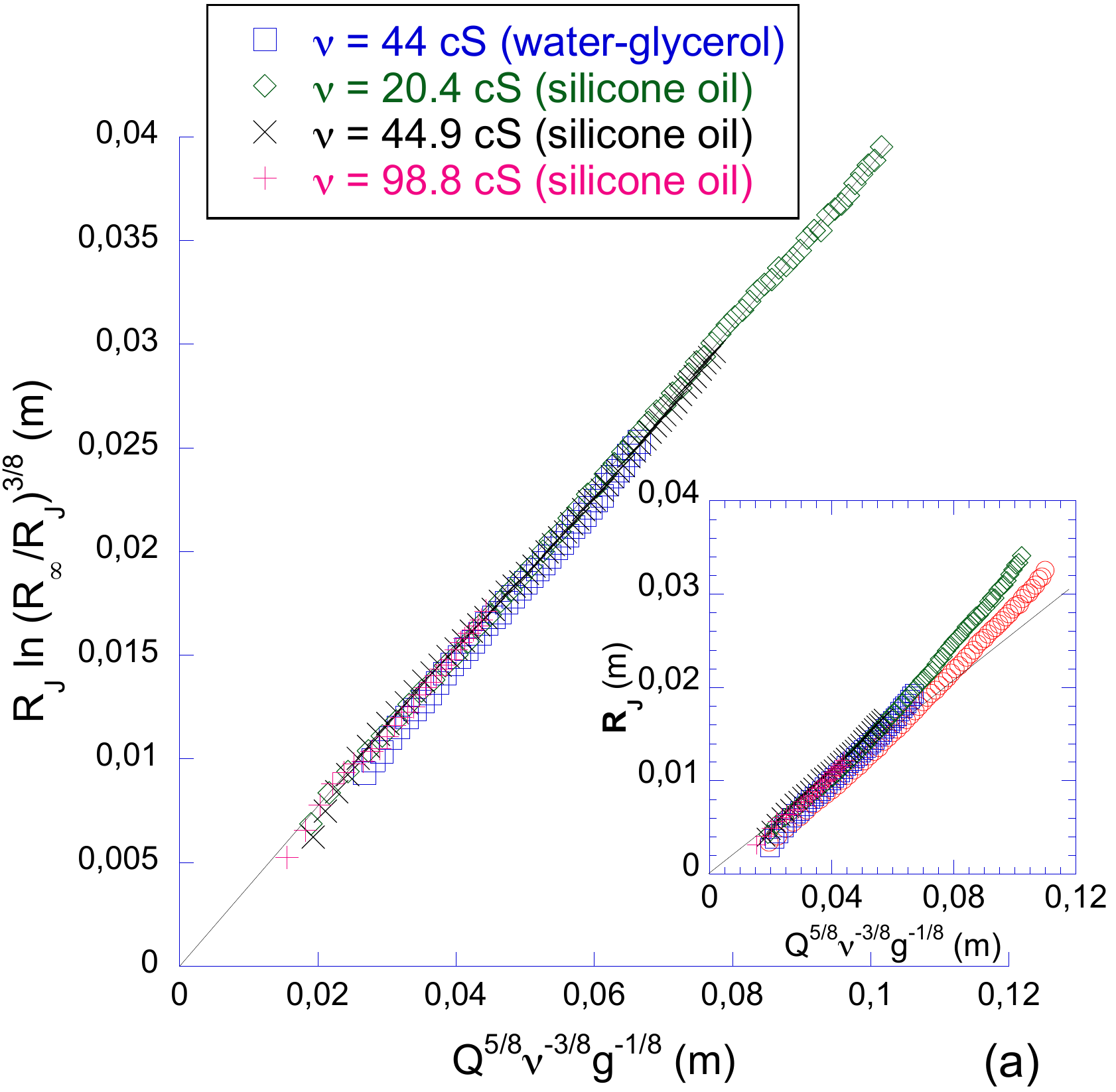}  \\
      \includegraphics[width=0.65 \columnwidth]{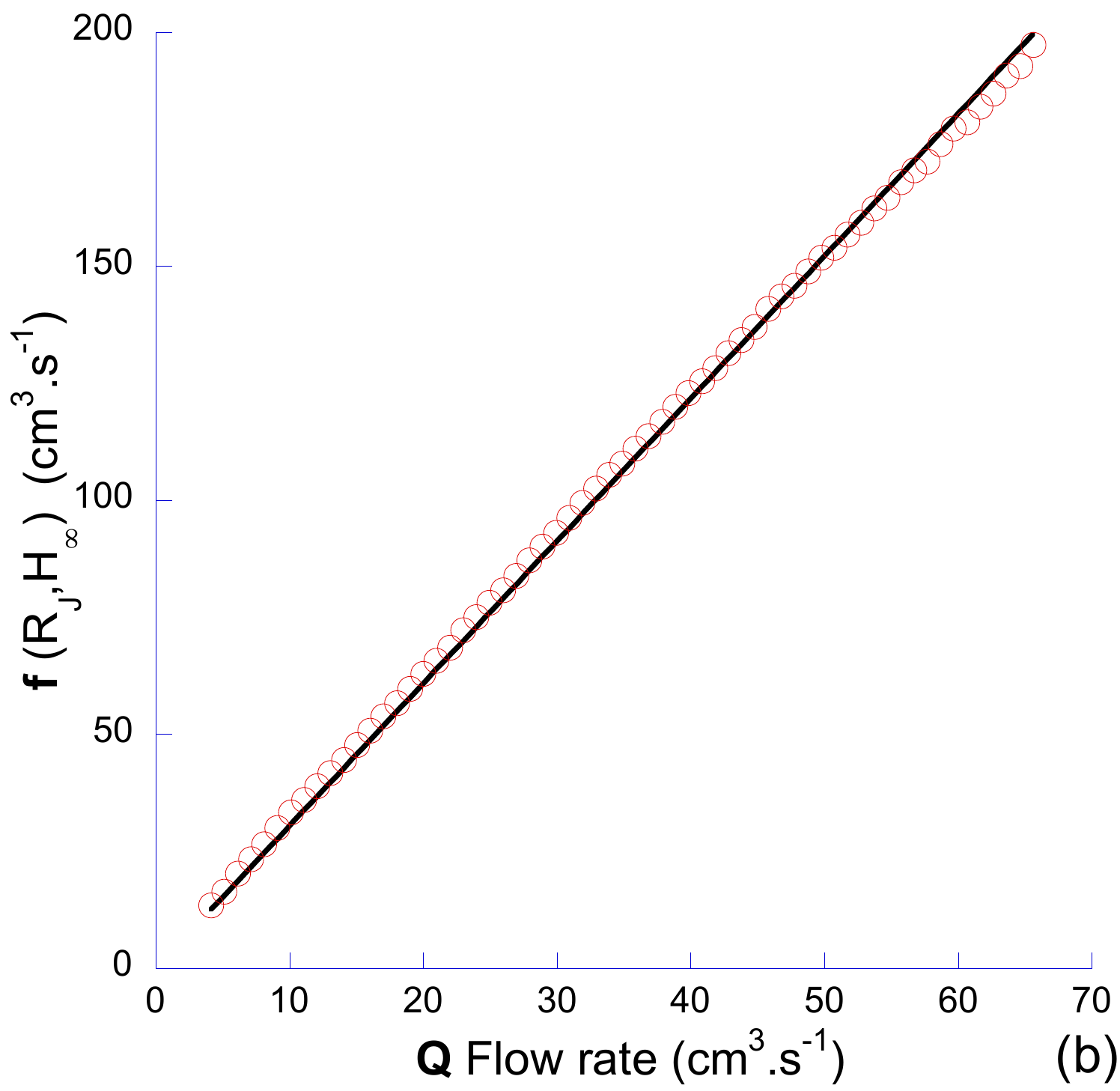}      
     \end{tabular}
\caption{\label{fig3} (a) Equation (\ref{e.4}) is tested experimentally for two different kind of liquids: silicone oil and water-glycerine mixture and for 4 different kinematic viscosities (20.4 cS, 44 cS, 44.9 cS and 98.8 cS). These curves, collapsing on a single master curve, are fitted by a linear function and a {constant} \masquer{universal} Froude number can be extract ( $Fr_2\approx 0.33$). Insert: \masquer{the radius of the jump $R_J$ is plotted versus the flow rate $Q$ in log-log for all the liquid presented in Fig.~\ref{fig3}. A law in $Q^{5/8}$ failed to describe accurately our data.}{the radius of the jump $R_J$ is plotted versus the Bohr \textit{et al.} scaling $Q^{5/8}\nu ^{-3/8} g^{-1/8}$ for all the liquid presented in Fig.~\ref{fig3}. Such a scaling law clearly failed to describe accurately our data.} (b) For lower viscosities of water glycerine mixture $H_\infty$ has to be taken into account: equation (\ref{e.4}) is plotted for water glycerine mixture with a kinematic viscosity of 18 cS. We observed that $H_\infty \simeq l_c \simeq 3$ mm. A linear function is also observed and the pre factor leads to a $Fr_2\approx 0.33$ which is consistent with the other results reported on Fig.~\ref{fig3}(a).}
\end{figure}
	\indent (i) In the partial wetting case (water-glycerol mixtures), the liquid escapes from the disk perimeter at several points under the shape of rivulets (the number of rivulets depends on the flow rate $Q$ and on the kinematic viscosity $\nu$, but is typically around 20). $H_\infty$ is then not fixed by the flow itself, but rather by some matching with a static contact line at the edge of the plate. We observed that $H_\infty$ appeared to be nearly constant and approximately equal to the capillary length $l_c\sim \sqrt{\frac{\gamma}{\rho g}}\sim 3$ mm.\\
	\indent (ii) In the total wetting case (silicone oils), the liquid wets the lateral edges of the disk with a flowing liquid layer. In this situation $H_\infty$ appears to depend on the flow rate, following a non trivial but weak power law and remaining of the same order of magnitude than the silicone oil capillary length $l_c\sim 1.5$ mm. In these conditions we can neglect $H_\infty ^4\ll \frac{6}{\pi} \frac{\nu Q}{g} ln (\frac{R_\infty}{R_J})$ and so we can consider that $H_J \simeq \frac{6}{\pi} \frac{\nu Q}{g} ln (\frac{R_\infty}{R_J})$. \\
	\indent Theses conditions are verified in the total wetting case for our range of flow rate (5-60 $\mathrm{cm^3.s^{-1}}$) and for every viscosities we used. In the partial wetting case the approximation is still good for the higher viscosity ($45$ cS) but become questionable for lower viscosities. \\
	\indent Writing that $Fr_J=U_J/\sqrt{gH_J}$ coupling to the flow rate conservation leads to : $Q^2=Fr^2 g 4 \pi ^2 R_J^2 H_J^3$. In the cases where we cannot neglect $H_\infty$, we obtain :
\begin{equation}
		\label{e.5}
Q=Fr \, f (R_J,H_\infty), 
\end{equation}	
with $f(R_J, H_\infty)= \sqrt{g} 2 \pi  R_J  (H_\infty^4+\frac{6}{\pi} \frac{\nu Q}{g} ln (\frac{R_\infty}{R_J}))^{3/8}$.\\
	\indent In the limit where the approximation of a negligible $H_\infty^4$ is still reasonable the previous equation can be re-written as: $Q^2=Fr^2 g 4 \pi ^2 R_J^2  (\frac{6}{\pi} \frac{\nu Q}{g} ln (\frac{R_\infty}{R_J}))^{3/4}$, and finally: 
\begin{equation}
		\label{e.4}
R_J (ln (\frac{R_\infty}{R_J}))^{3/8}= \beta Fr^{-1} Q^{5/8}\nu ^{-3/8} g^{-1/8}, 
\end{equation}
where $\beta = (\frac{6}{\pi})^{-3/8}\frac{1}{2\pi }$. One can notice here, that we recover Bohr scaling but with a non negligible logarithmic pre factor. Experimental results are compared to these laws in Fig.~\ref{fig3}(a) and  Fig.~\ref{fig3}(b). From these data two significant lessons may be learned: \\
	\indent (i) At first we can conclude that our description provides an analytical law in excellent agreement with \masquer{the experience} {experiments} in all the tested cases {and} with the same jump exit Froude number, as expected from the matching condition. Particularly we can observe that the effect of the viscosity is well described by the $-3/8$ exponent. More precisely, the value of the Froude number that can be extracted from all the set of data (Fig.~\ref{fig3}(a) and  Fig.~\ref{fig3}(b)), is the one expected: $Fr_2=0.33$. {The logarithmic term and the influence of $H_\infty$ in (\ref{e.5}) appears to be important: the scaling proposed by Bohr \textit{et al.} is tested on the insert of Fig.~\ref{fig3}(a) for all our data set and shows clearly that the scaling failed to predict accurately the dependance of $R_J$ on the different physical parameters.}\\
	\indent (ii) As one can observe, the surface tension, which is the main difference between silicone oils and water-glycerol mixtures, is absent in equations (\ref{e.5}) and  (\ref{e.4}). So one can conclude that the role of surface tension is negligible for large enough radius. Indeed, according to Bush \textit{et al.}, surface tension plays a minor role in the momentum conservation at the jump level \cite{Bush_2003, Kasimov_2008} that reads: \\
	\begin{equation}
			\label{e.7}
\frac{1}{2}g(H_J^2-h_J^2)+\frac {\gamma}{\rho} \frac{H_J-h_J}{R_J}=\int _0^{h_J} u^2  \mathrm{d}z - \int _0^{H_J} U^2  \mathrm{d}z ,
\end{equation}	
	where $h_J$, $u$, $H_J$ and $U$ are respectively the height and the liquid velocity at the either side of the jump and $\rho$ the liquid density.\\
	\indent A scaling analysis gives an estimate \masquer{value} for a jump radius where the surface tension term has the same weight than the others : $R_J\sim \frac {2\gamma}{\rho g (H_J+h_J)}\sim10^{-3}$ m (calculate for silicone oil). So we recover that for large enough radii the surface tension becomes negligible.\\
		\indent Our result of a fixed Froude number has numerous consequences, particularly about the inner zone of the hydraulic jump. Using the momentum conservation (cf. equation (\ref{e.7})) one can predict that the Froude number just before the jump is also constant (except for low radii when surface tension has to be taken into account). In this way, the average speed before the jump $u$ can also be estimated. This average speed $u$ appears to be weakly dependent on the flow rate ($u\sim Q^{0.05}$). This result is perfectly coherent with a previous observation reported by the authors \cite {Duchesne_2013} that the flow structure at the jump appears to be quite independent of the flow rate $Q$.\\
	\indent To sum up our result, a \masquer{universal}{constant} Froude number $Fr$ \masquer{(}independent of the flow rate, viscosity and surface tension\masquer{)} is observed in the case of a circular hydraulic jump without any confinement wall \masquer{and seems to be locked to a constant value}. A lubrication description of the outer liquid thickness profile has been proposed and is well confirmed by experimental data. This profile combined with the {locking-}condition on the Froude number leads to a law linking the radius of the jump and the experimental conditions: $R_J (ln (\frac{R_\infty}{R_J}))^{3/8} \sim Q^{5/8}\nu ^{-3/8} g^{-1/8}$ valid for a large range of experimental conditions. This law was experimentally tested and our data show a good agreement with our mathematical approach. This law recovers the scaling expected by Bohr \textit{et al.} but with a non negligible logarithmic correction. \masquer{While the reason for a fixed Froude number is still unknown, our approach indicates that it is related to the inner flow, before the jump. Further studies on this subject are under way.}\\
	\indent {The origin of a constant Froude number is still unknown. We have explored its possible dependance upon the geometrical parameters (nozzle diameters, disk diameters, nozzle to disk distance). It is independant of the nozzle to disk distance (1 \% of variation for a change between 4 and 40 mm) and slightly dependent of the nozzle diameter (10 \% for a  variation from 1.1 to 7 mm) and of the disk radius (5 \% for a  variation from $R_{\infty}=10$ cm to $R_{\infty}=15$ cm). However we emphasize here that in all the tested cases we found a constant Froude number and an excellent agreement between the jump radius and the laws (\ref{e.5}) and (\ref{e.4}) observed above.}   \\
	\acknowledgements{Acknowledgements. We thank Andrew Belmonte for carefully reading the manuscript \masquer{and for his valuable comments} {and Henri Lhuissier for discussions and additional numerical simulations}. The authors also wish to thank Laurent Rhea, and Mathieu\masquer{x} Receveur for technical assistance. One of us (L. Limat) thanks Marc Durand for discussions on the outer flow problem. This work was sponsored by the French National Agency for Research (Freeflow project ref. ANR-11-B504-001-01).}
\bibliography{bibliojump.bib}

%merlin.mbs apsrev4-1.bst 2010-07-25 4.21a (PWD, AO, DPC) hacked
%Control: key (0)
%Control: author (8) initials jnrlst
%Control: editor formatted (1) identically to author
%Control: production of article title (-1) disabled
%Control: page (0) single
%Control: year (1) truncated
%Control: production of eprint (0) enabled
\begin{thebibliography}{24}%
\makeatletter
\providecommand \@ifxundefined [1]{%
 \@ifx{#1\undefined}
}%
\providecommand \@ifnum [1]{%
 \ifnum #1\expandafter \@firstoftwo
 \else \expandafter \@secondoftwo
 \fi
}%
\providecommand \@ifx [1]{%
 \ifx #1\expandafter \@firstoftwo
 \else \expandafter \@secondoftwo
 \fi
}%
\providecommand \natexlab [1]{#1}%
\providecommand \enquote  [1]{``#1''}%
\providecommand \bibnamefont  [1]{#1}%
\providecommand \bibfnamefont [1]{#1}%
\providecommand \citenamefont [1]{#1}%
\providecommand \href@noop [0]{\@secondoftwo}%
\providecommand \href [0]{\begingroup \@sanitize@url \@href}%
\providecommand \@href[1]{\@@startlink{#1}\@@href}%
\providecommand \@@href[1]{\endgroup#1\@@endlink}%
\providecommand \@sanitize@url [0]{\catcode `\\12\catcode `\$12\catcode
  `\&12\catcode `\#12\catcode `\^12\catcode `\_12\catcode `\%12\relax}%
\providecommand \@@startlink[1]{}%
\providecommand \@@endlink[0]{}%
\providecommand \url  [0]{\begingroup\@sanitize@url \@url }%
\providecommand \@url [1]{\endgroup\@href {#1}{\urlprefix }}%
\providecommand \urlprefix  [0]{URL }%
\providecommand \Eprint [0]{\href }%
\providecommand \doibase [0]{http://dx.doi.org/}%
\providecommand \selectlanguage [0]{\@gobble}%
\providecommand \bibinfo  [0]{\@secondoftwo}%
\providecommand \bibfield  [0]{\@secondoftwo}%
\providecommand \translation [1]{[#1]}%
\providecommand \BibitemOpen [0]{}%
\providecommand \bibitemStop [0]{}%
\providecommand \bibitemNoStop [0]{.\EOS\space}%
\providecommand \EOS [0]{\spacefactor3000\relax}%
\providecommand \BibitemShut  [1]{\csname bibitem#1\endcsname}%
\let\auto@bib@innerbib\@empty
%</preamble>
\bibitem [{\citenamefont {Chanson}(2009)}]{chanson_2009}%
  \BibitemOpen
  \bibfield  {author} {\bibinfo {author} {\bibfnamefont {H.}~\bibnamefont
  {Chanson}},\ }\href@noop {} {\bibfield  {journal} {\bibinfo  {journal} {Eur.
  J. Mech. B: Fluids}\ }\textbf {\bibinfo {volume} {28}},\ \bibinfo {pages}
  {191} (\bibinfo {year} {2009})}\BibitemShut {NoStop}%
\bibitem [{\citenamefont {Lighthill}(1978)}]{Lighthill_78}%
  \BibitemOpen
  \bibfield  {author} {\bibinfo {author} {\bibfnamefont {J.}~\bibnamefont
  {Lighthill}},\ }\href@noop {} {\emph {\bibinfo {title} {Waves In Fluids}}}\
  (\bibinfo  {publisher} {Cambridge University Press},\ \bibinfo {year}
  {1978})\BibitemShut {NoStop}%
\bibitem [{\citenamefont {Chanson}(2012)}]{chanson_mascaret}%
  \BibitemOpen
  \bibfield  {author} {\bibinfo {author} {\bibfnamefont {H.}~\bibnamefont
  {Chanson}},\ }\href@noop {} {\emph {\bibinfo {title} {Tidal bores, aegir,
  eagre, mascaret, pororoca: Theory and observations}}}\ (\bibinfo  {publisher}
  {World Scientific},\ \bibinfo {year} {2012})\BibitemShut {NoStop}%
\bibitem [{\citenamefont {Jannes}\ \emph {et~al.}(2011)\citenamefont {Jannes},
  \citenamefont {Piquet}, \citenamefont {Maissa}, \citenamefont {Mathis},\ and\
  \citenamefont {Rousseaux}}]{Rousseaux_2011}%
  \BibitemOpen
  \bibfield  {author} {\bibinfo {author} {\bibfnamefont {G.}~\bibnamefont
  {Jannes}}, \bibinfo {author} {\bibfnamefont {R.}~\bibnamefont {Piquet}},
  \bibinfo {author} {\bibfnamefont {P.}~\bibnamefont {Maissa}}, \bibinfo
  {author} {\bibfnamefont {C.}~\bibnamefont {Mathis}}, \ and\ \bibinfo {author}
  {\bibfnamefont {G.}~\bibnamefont {Rousseaux}},\ }\href@noop {} {\bibfield
  {journal} {\bibinfo  {journal} {Physical Review E}\ }\textbf {\bibinfo
  {volume} {83}},\ \bibinfo {pages} {056312} (\bibinfo {year}
  {2011})}\BibitemShut {NoStop}%
\bibitem [{\citenamefont {Weinfurtner}\ \emph {et~al.}(2011)\citenamefont
  {Weinfurtner}, \citenamefont {Tedford}, \citenamefont {Penrice},
  \citenamefont {Unruh},\ and\ \citenamefont {Lawrence}}]{Weinfurtner_2011}%
  \BibitemOpen
  \bibfield  {author} {\bibinfo {author} {\bibfnamefont {S.}~\bibnamefont
  {Weinfurtner}}, \bibinfo {author} {\bibfnamefont {E.~W.}\ \bibnamefont
  {Tedford}}, \bibinfo {author} {\bibfnamefont {M.~C.~J.}\ \bibnamefont
  {Penrice}}, \bibinfo {author} {\bibfnamefont {W.~G.}\ \bibnamefont {Unruh}},
  \ and\ \bibinfo {author} {\bibfnamefont {G.~A.}\ \bibnamefont {Lawrence}},\
  }\href {\doibase 10.1103/PhysRevLett.106.021302} {\bibfield  {journal}
  {\bibinfo  {journal} {Phys. Rev. Lett.}\ }\textbf {\bibinfo {volume} {106}},\
  \bibinfo {pages} {021302} (\bibinfo {year} {2011})}\BibitemShut {NoStop}%
\bibitem [{\citenamefont {Belanger}(1828)}]{belanger}%
  \BibitemOpen
  \bibfield  {author} {\bibinfo {author} {\bibfnamefont {J.~B. C.~J.}\
  \bibnamefont {Belanger}},\ }\href@noop {} {\emph {\bibinfo {title} {Essai sur
  la Solution Num{\'e}rique de quelques Probl{\`e}mes Relatifs au Mouvement
  Permanent des Eaux Courantes}}}\ (\bibinfo  {publisher} {Carilian-Goeury
  (Paris)},\ \bibinfo {year} {1828})\BibitemShut {NoStop}%
\bibitem [{\citenamefont {Bohr}\ \emph {et~al.}(1996)\citenamefont {Bohr},
  \citenamefont {Ellegaard}, \citenamefont {Hansen},\ and\ \citenamefont
  {Haaning}}]{Bohr_96}%
  \BibitemOpen
  \bibfield  {author} {\bibinfo {author} {\bibfnamefont {T.}~\bibnamefont
  {Bohr}}, \bibinfo {author} {\bibfnamefont {C.}~\bibnamefont {Ellegaard}},
  \bibinfo {author} {\bibfnamefont {A.~E.}\ \bibnamefont {Hansen}}, \ and\
  \bibinfo {author} {\bibfnamefont {A.}~\bibnamefont {Haaning}},\ }\href@noop
  {} {\bibfield  {journal} {\bibinfo  {journal} {Physica B}\ }\textbf {\bibinfo
  {volume} {228}},\ \bibinfo {pages} {1} (\bibinfo {year} {1996})}\BibitemShut
  {NoStop}%
\bibitem [{\citenamefont {Bush}\ \emph {et~al.}(2006)\citenamefont {Bush},
  \citenamefont {Aristoff},\ and\ \citenamefont {Hosoi}}]{Bush_2006}%
  \BibitemOpen
  \bibfield  {author} {\bibinfo {author} {\bibfnamefont {J.~W.~M.}\
  \bibnamefont {Bush}}, \bibinfo {author} {\bibfnamefont {J.~M.}\ \bibnamefont
  {Aristoff}}, \ and\ \bibinfo {author} {\bibfnamefont {A.~E.}\ \bibnamefont
  {Hosoi}},\ }\href@noop {} {\bibfield  {journal} {\bibinfo  {journal} {J.
  Fluid Mech.}\ }\textbf {\bibinfo {volume} {558}},\ \bibinfo {pages} {33}
  (\bibinfo {year} {2006})}\BibitemShut {NoStop}%
\bibitem [{\citenamefont {Martens}\ \emph {et~al.}(2012)\citenamefont
  {Martens}, \citenamefont {Watanabe},\ and\ \citenamefont {Bohr}}]{bohr_12}%
  \BibitemOpen
  \bibfield  {author} {\bibinfo {author} {\bibfnamefont {E.~A.}\ \bibnamefont
  {Martens}}, \bibinfo {author} {\bibfnamefont {S.}~\bibnamefont {Watanabe}}, \
  and\ \bibinfo {author} {\bibfnamefont {T.}~\bibnamefont {Bohr}},\ }\href@noop
  {} {\bibfield  {journal} {\bibinfo  {journal} {Physical Review E}\ }\textbf
  {\bibinfo {volume} {85}},\ \bibinfo {pages} {036316} (\bibinfo {year}
  {2012})}\BibitemShut {NoStop}%
\bibitem [{\citenamefont {Rayleigh}(1914)}]{Rayleigh_1914}%
  \BibitemOpen
  \bibfield  {author} {\bibinfo {author} {\bibfnamefont {L.}~\bibnamefont
  {Rayleigh}},\ }\href@noop {} {\bibfield  {journal} {\bibinfo  {journal}
  {Proc. R. Soc. London A}\ }\textbf {\bibinfo {volume} {90}},\ \bibinfo
  {pages} {324} (\bibinfo {year} {1914})}\BibitemShut {NoStop}%
\bibitem [{\citenamefont {Tani}(1949)}]{Tani_48}%
  \BibitemOpen
  \bibfield  {author} {\bibinfo {author} {\bibfnamefont {I.}~\bibnamefont
  {Tani}},\ }\href@noop {} {\bibfield  {journal} {\bibinfo  {journal} {J. Phys.
  Soc. Japan}\ }\textbf {\bibinfo {volume} {4}},\ \bibinfo {pages} {212}
  (\bibinfo {year} {1949})}\BibitemShut {NoStop}%
\bibitem [{\citenamefont {Watson}(1964)}]{Watson_JFM}%
  \BibitemOpen
  \bibfield  {author} {\bibinfo {author} {\bibfnamefont {E.~J.}\ \bibnamefont
  {Watson}},\ }\href@noop {} {\bibfield  {journal} {\bibinfo  {journal} {J.
  Fluid Mech.}\ }\textbf {\bibinfo {volume} {20}},\ \bibinfo {pages} {481}
  (\bibinfo {year} {1964})}\BibitemShut {NoStop}%
\bibitem [{\citenamefont {Craik}\ \emph {et~al.}(1981)\citenamefont {Craik},
  \citenamefont {Latham}, \citenamefont {Fawkes},\ and\ \citenamefont
  {Gribbon}}]{Craik_81}%
  \BibitemOpen
  \bibfield  {author} {\bibinfo {author} {\bibfnamefont {A.~D.~D.}\
  \bibnamefont {Craik}}, \bibinfo {author} {\bibfnamefont {R.~C.}\ \bibnamefont
  {Latham}}, \bibinfo {author} {\bibfnamefont {M.~J.}\ \bibnamefont {Fawkes}},
  \ and\ \bibinfo {author} {\bibfnamefont {P.~W.~F.}\ \bibnamefont {Gribbon}},\
  }\href@noop {} {\bibfield  {journal} {\bibinfo  {journal} {J. Fluid Mech.}\
  }\textbf {\bibinfo {volume} {112}},\ \bibinfo {pages} {347} (\bibinfo {year}
  {1981})}\BibitemShut {NoStop}%
\bibitem [{\citenamefont {Bohr}\ \emph {et~al.}(1993)\citenamefont {Bohr},
  \citenamefont {Dimon},\ and\ \citenamefont {Putkaradze}}]{Bohr_1993}%
  \BibitemOpen
  \bibfield  {author} {\bibinfo {author} {\bibfnamefont {T.}~\bibnamefont
  {Bohr}}, \bibinfo {author} {\bibfnamefont {P.}~\bibnamefont {Dimon}}, \ and\
  \bibinfo {author} {\bibfnamefont {V.}~\bibnamefont {Putkaradze}},\
  }\href@noop {} {\bibfield  {journal} {\bibinfo  {journal} {J. Fluid Mech.}\
  }\textbf {\bibinfo {volume} {254}},\ \bibinfo {pages} {635} (\bibinfo {year}
  {1993})}\BibitemShut {NoStop}%
\bibitem [{\citenamefont {Higuera}(1994)}]{higuera_1994}%
  \BibitemOpen
  \bibfield  {author} {\bibinfo {author} {\bibfnamefont {F.}~\bibnamefont
  {Higuera}},\ }\href@noop {} {\bibfield  {journal} {\bibinfo  {journal}
  {Journal of fluid Mechanics}\ }\textbf {\bibinfo {volume} {274}},\ \bibinfo
  {pages} {69} (\bibinfo {year} {1994})}\BibitemShut {NoStop}%
\bibitem [{\citenamefont {Bush}\ and\ \citenamefont
  {Aristoff}(2003)}]{Bush_2003}%
  \BibitemOpen
  \bibfield  {author} {\bibinfo {author} {\bibfnamefont {J.~W.~M.}\
  \bibnamefont {Bush}}\ and\ \bibinfo {author} {\bibfnamefont {J.~M.}\
  \bibnamefont {Aristoff}},\ }\href@noop {} {\bibfield  {journal} {\bibinfo
  {journal} {J. Fluid Mech.}\ }\textbf {\bibinfo {volume} {489}},\ \bibinfo
  {pages} {229} (\bibinfo {year} {2003})}\BibitemShut {NoStop}%
\bibitem [{\citenamefont {Yokoi}\ and\ \citenamefont {Xiao}(2000)}]{yokoi2000}%
  \BibitemOpen
  \bibfield  {author} {\bibinfo {author} {\bibfnamefont {K.}~\bibnamefont
  {Yokoi}}\ and\ \bibinfo {author} {\bibfnamefont {F.}~\bibnamefont {Xiao}},\
  }\href@noop {} {\bibfield  {journal} {\bibinfo  {journal} {Physical Review
  E}\ }\textbf {\bibinfo {volume} {61}},\ \bibinfo {pages} {R1016} (\bibinfo
  {year} {2000})}\BibitemShut {NoStop}%
\bibitem [{\citenamefont {Dressaire}\ \emph {et~al.}(2009)\citenamefont
  {Dressaire}, \citenamefont {Courbin}, \citenamefont {Crest},\ and\
  \citenamefont {Stone}}]{Stone_2009}%
  \BibitemOpen
  \bibfield  {author} {\bibinfo {author} {\bibfnamefont {E.}~\bibnamefont
  {Dressaire}}, \bibinfo {author} {\bibfnamefont {L.}~\bibnamefont {Courbin}},
  \bibinfo {author} {\bibfnamefont {J.}~\bibnamefont {Crest}}, \ and\ \bibinfo
  {author} {\bibfnamefont {H.~A.}\ \bibnamefont {Stone}},\ }\href@noop {}
  {\bibfield  {journal} {\bibinfo  {journal} {Phys. Rev. Lett.}\ }\textbf
  {\bibinfo {volume} {102}} (\bibinfo {year} {2009})}\BibitemShut {NoStop}%
\bibitem [{\citenamefont {Dasgupta}\ and\ \citenamefont
  {Govindarajan}(2010)}]{dasgupta2010}%
  \BibitemOpen
  \bibfield  {author} {\bibinfo {author} {\bibfnamefont {R.}~\bibnamefont
  {Dasgupta}}\ and\ \bibinfo {author} {\bibfnamefont {R.}~\bibnamefont
  {Govindarajan}},\ }\href@noop {} {\bibfield  {journal} {\bibinfo  {journal}
  {Physics of Fluids}\ }\textbf {\bibinfo {volume} {22}},\ \bibinfo {pages}
  {112108} (\bibinfo {year} {2010})}\BibitemShut {NoStop}%
\bibitem [{\citenamefont {Maynes}\ \emph {et~al.}(2011)\citenamefont {Maynes},
  \citenamefont {Johnson},\ and\ \citenamefont {Webb}}]{Maynes_2011}%
  \BibitemOpen
  \bibfield  {author} {\bibinfo {author} {\bibfnamefont {D.}~\bibnamefont
  {Maynes}}, \bibinfo {author} {\bibfnamefont {M.}~\bibnamefont {Johnson}}, \
  and\ \bibinfo {author} {\bibfnamefont {B.~W.}\ \bibnamefont {Webb}},\
  }\href@noop {} {\bibfield  {journal} {\bibinfo  {journal} {Phys. Fluids}\
  }\textbf {\bibinfo {volume} {23}} (\bibinfo {year} {2011})}\BibitemShut
  {NoStop}%
\bibitem [{\citenamefont {Rojas}\ \emph {et~al.}(2013)\citenamefont {Rojas},
  \citenamefont {Argentina},\ and\ \citenamefont {Tirapegui}}]{rojas_2013}%
  \BibitemOpen
  \bibfield  {author} {\bibinfo {author} {\bibfnamefont {N.}~\bibnamefont
  {Rojas}}, \bibinfo {author} {\bibfnamefont {M.}~\bibnamefont {Argentina}}, \
  and\ \bibinfo {author} {\bibfnamefont {E.}~\bibnamefont {Tirapegui}},\
  }\href@noop {} {\bibfield  {journal} {\bibinfo  {journal} {Physics of Fluids
  (1994-present)}\ }\textbf {\bibinfo {volume} {25}},\ \bibinfo {pages}
  {042105} (\bibinfo {year} {2013})}\BibitemShut {NoStop}%
\bibitem [{\citenamefont {Watanabe}\ \emph {et~al.}(2003)\citenamefont
  {Watanabe}, \citenamefont {Putkaradze},\ and\ \citenamefont
  {Bohr}}]{bohr_2003}%
  \BibitemOpen
  \bibfield  {author} {\bibinfo {author} {\bibfnamefont {S.}~\bibnamefont
  {Watanabe}}, \bibinfo {author} {\bibfnamefont {V.}~\bibnamefont
  {Putkaradze}}, \ and\ \bibinfo {author} {\bibfnamefont {T.}~\bibnamefont
  {Bohr}},\ }\href@noop {} {\bibfield  {journal} {\bibinfo  {journal} {Journal
  of Fluid Mechanics}\ }\textbf {\bibinfo {volume} {480}},\ \bibinfo {pages}
  {233} (\bibinfo {year} {2003})}\BibitemShut {NoStop}%
\bibitem [{\citenamefont {Kasimov}(2008)}]{Kasimov_2008}%
  \BibitemOpen
  \bibfield  {author} {\bibinfo {author} {\bibfnamefont {A.~R.}\ \bibnamefont
  {Kasimov}},\ }\href@noop {} {\bibfield  {journal} {\bibinfo  {journal} {J.
  Fluid Mech.}\ }\textbf {\bibinfo {volume} {601}},\ \bibinfo {pages} {189}
  (\bibinfo {year} {2008})}\BibitemShut {NoStop}%
\bibitem [{\citenamefont {Duchesne}\ \emph {et~al.}(2013)\citenamefont
  {Duchesne}, \citenamefont {Savaro}, \citenamefont {Lebon}, \citenamefont
  {Pirat},\ and\ \citenamefont {Limat}}]{Duchesne_2013}%
  \BibitemOpen
  \bibfield  {author} {\bibinfo {author} {\bibfnamefont {A.}~\bibnamefont
  {Duchesne}}, \bibinfo {author} {\bibfnamefont {C.}~\bibnamefont {Savaro}},
  \bibinfo {author} {\bibfnamefont {L.}~\bibnamefont {Lebon}}, \bibinfo
  {author} {\bibfnamefont {C.}~\bibnamefont {Pirat}}, \ and\ \bibinfo {author}
  {\bibfnamefont {L.}~\bibnamefont {Limat}},\ }\href@noop {} {\bibfield
  {journal} {\bibinfo  {journal} {Europhys. Lett.}\ }\textbf {\bibinfo {volume}
  {102}} (\bibinfo {year} {2013})}\BibitemShut {NoStop}%
\end{thebibliography}%
\end{document}